# Superconducting Dome and Quantum Criticality in Two-Dimensional NbO$_2$ Triangular Lattice

Takuto Soma[1,*], Kohei Yoshimatsu[1], and Akira Ohtomo[1]

[1]Department of Chemical Science and Engineering, Institute of Science Tokyo; Tokyo, 152-8552, Japan

**ABSTRACT** The emergence of superconductivity with strong correlation is one of the most attracted issues in condensed-matter physics, as seen in various unconventional superconductors. Here we show a new strongly correlated superconductor Li$_{1-x}$NbO$_2$ with rich characteristics such as two-dimensional, geometrically frustrated, and triangular NbO$_2$ lattice and correlated flat-band-like electronic states. We revealed the electronic phase diagram by implementing Li-ion electrochemical cells with LiNbO$_2$ epitaxial films. The Li-ion deintercalation increased the hole-doping level in NbO$_2$ layer, along which a band insulator LiNbO$_2$ underwent to a Fermi-liquid (FL) metal and superconductor associated with non-Fermi liquid (NFL) characters. The evolution of the NFL state coincided with the suppression of the Kondo-singlet formation near the superconducting dome, which linked superconductivity with quantum criticality. The obtained phase diagram involves general aspects of strongly correlated superconductors and bridges the gap between various systems.

## I. INTRODUCTION

Superconductivity (SC) associated with strong electron correlation has been regarded as one of the biggest problems in condensed-matter physics [1–4]. Such exotic classes of superconductors include high-$T_c$ cuprates [5], heavy-fermion superconductors [6], iron-based superconductors [7], recently found nickelates [8,9], and Moiré superlattices in two-dimensional (2D) materials [10]. In these compounds, SC usually appears with the suppression of magnetisms around a quantum critical point (QCP) with strong fluctuation. Therefore, magnetic interaction and quantum critical fluctuation are thought to be an important pairing factor to create exotic superconducting states [1,2]. In this vein, correlated narrow bands, as represented by Jahn-Teller distortion-induced Cu 3$d_{x^2-y^2}$ single bands in cuprates [1,11] and Moiré bands in twisted 2D materials [4,11], play important roles to quench electron kinetic energy and realize the unconventional SC. Similarly, frustrated Kagome compounds with flat bands have recently attracted much attention [12–14]. Despite the variety of material classes, the small number of strongly correlated superconductors prevents us from investigating common principles covering all or a part of classes [1–4].

In this Letter, we focus on a layered niobite LiNbO$_2$, which consists of alternately stacked NbO$_2$ and Li layers [Fig. 1(a)] [15]. LiNbO$_2$ is a band insulator, while SC appears around 5 K in Li-deficient Li$_{0.45}$NbO$_2$ [16], which can be understood to be analogous to cuprates, namely hole doping to a 2D NbO$_2$ superconducting layer from a Li$^+$ charge reservoir layer. Not only the 2D layer in the *ab* plane, but also a 1D tunnel along the *c*-axis in Li$_{1-x}$NbO$_2$ are suitable for electrochemical Li-ion (de)intercalation [Fig. 1(b)] [17]. Furthermore, Li$_0$NbO$_2$ is the only isostructural oxide to 2*H*-type transition-metal dichalcogenides, and thus Li$_{1-x}$NbO$_2$ are regarded as Li-intercalated 2*H*-type NbO$_2$ [18,19]. Since Nb$^{4+}$ (4$d^1$) ions in the 2*H*-type NbO$_2$ form the 2D triangular lattice with $S = 1/2$, strong spin frustration is expected in Li$_{1-x}$NbO$_2$. In addition, its band structure displays suitable characters as a host of strongly correlated electrons, as schematically illustrated in Fig. 1(c) [20–22]. A triangular prismatic coordination with strong ligand field of O$^{2-}$ makes a highly isolated Nb 4$d_{z^2}$ single band, as verified by the DFT calculation [Fig. 1(d)] and photoemission spectroscopy [22]. Narrow width of the Nb 4$d_{z^2}$ band highlights the importance of electron correlation. In previous studies, for example, the visible-light transparency can be rather enhanced upon the hole doping in superconducting Li$_{1-x}$NbO$_2$ [23,24]. More importantly, Li deintercalation makes the Fermi level ($E_F$) in the Nb 4$d_{z^2}$ band deeper. The $E_F$ eventually reaches the flat-band-like small dispersed states around −1 eV [Fig. 1(d)], with the nodal surface states in $k_z = \pi$ [Fig. 1(e)] [25–28]. In other words, SC in Li$_{1-x}$NbO$_2$ is triggered by the heavy hole doping that make $E_F$ reachable within small dispersion bands where the kinetic energy of the electrons is quenched. This mechanism suggests strong unconventionality in the emergence of the SC [3,4].

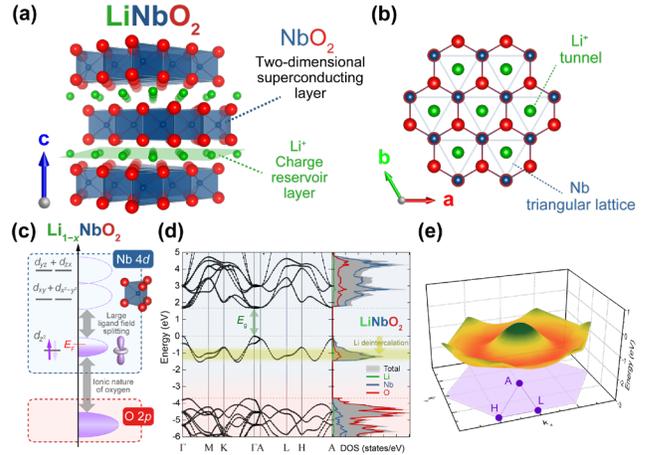

FIG. 1. (a) Schematics of the layered structure for LiNbO$_2$ (cross-sectional view) and (b) two-dimensional NbO$_2$ layers (top view). (c) Band structure of LiNbO$_2$ obtained from DFT calculation (18,20–22). (d) Schematics of band diagram for Li$_{1-x}$NbO$_2$. Correlated narrow Nb 4$d_{z^2}$ single-band structure is attributed to a strong ligand field of oxygen with the triangular prism coordination. (e) Contour plot of Nb 4$d_{z^2}$ band dispersion in the AHL plane. High symmetry points of hexagonal Brillouin zone are shown in the bottom.

*Contact author: soma@mct.isct.ac.jp

Regardless of the interesting features of Li$_{1-x}$NbO$_2$, its electronic states are hardly explored because Li-deficient phases are metastable at ambient conditions. The lack of highly crystalline and homogeneous samples prevents us from investigating intrinsic properties [16,29]. To overcome these problems, we have developed methods to obtain epitaxial films of superconducting Li$_{1-x}$NbO$_2$ [23], perform the uniform deintercalation reaction, and measure transport properties *in-situ* [30]. Moreover, combining electrochemical cells with epitaxial films, we are able to control carrier concentration precisely and measure intrinsic physical properties of Li$_{1-x}$NbO$_2$ without undesired variations among samples [30]. Here we establish the electronic phase diagram of Li$_{1-x}$NbO$_2$ to elucidate a superconducting dome and quantum critical behaviors in the frustrated 2D NbO$_2$ triangular lattice.

## II. METHODS
The experimental methods are described in the Supplemental Material [31].

## III. RESULTS
### A. Electrochemical cell for controlling carrier concentration of Li$_{1-x}$NbO$_2$

Figure 2(a) shows schematics of the electrochemical cell used in this study. It enabled us to investigate the systematic evolution of electronic states in a Li$_{1-x}$NbO$_2$ single film. Depending on potentials applied to the cell by using a potentiostat, carrier concentration in Li$_{1-x}$NbO$_2$ was systematically and widely modulated. The carrier concentrations, evaluated as $1/eR_H$ from Hall coefficients $R_H$ measured at 100 K [Fig. S1(a) [31]], showed systematic variations against charges and/or currents fed into the electrochemical cell (Fig. S1(b) [31]), which was supported by the following electrochemical reaction:

$$\text{LiNbO}_2 \rightleftarrows \text{Li}_{1-x}\text{NbO}_2 + x\text{Li}^+ + e^-. \quad (1)$$

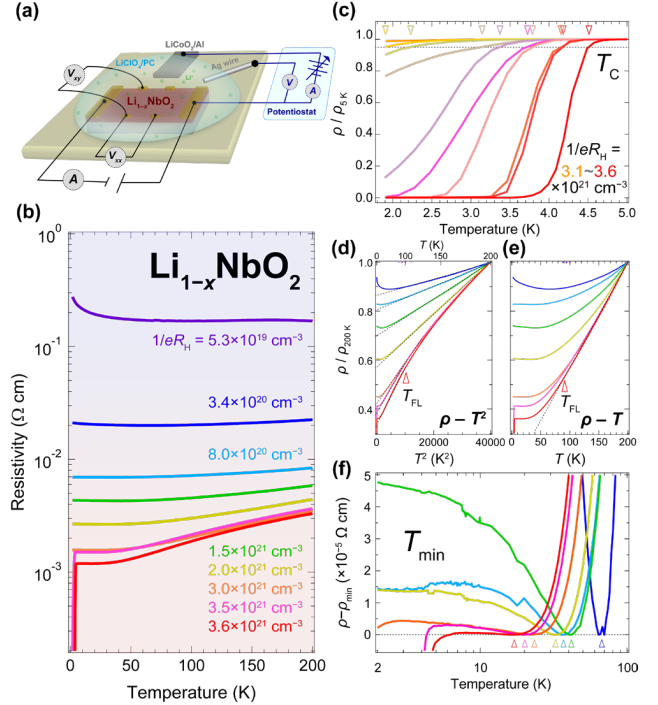

FIG. 2. (a) Schematics of the electrochemical cell used in this study. (b-f) Temperature ($T$)-dependent resistivity ($\rho$) for Li$_{1-x}$NbO$_2$ films with various carrier concentrations. (b) Wide temperature region. (c) Magnification (with curves normalized by those at 5 K) near superconducting transitions ($T_c$ are indicated by triangular symbols pointing intersections with the dotted line). (d) $T^2$- and (e) $T$-dependent $\rho$ normalized by those at 200 K. Dashed lines in (d) and (e) indicate results of linear fits. (f) Logarithmic $T$-dependent $\rho-\rho_{min}$ where triangular symbols indicate $\rho_{min}$.

The maximum accessible hole-carrier concentration was $1/eR_H = 3.6 \times 10^{21}$ cm$^{-3}$ in this study, which was limited by the crystallographic instability at $x > 0.55$ as reported for bulk Li$_{1-x}$NbO$_2$ [16,17,29]. In fact, the interlayer distance [23] and Li 1$s$ XPS spectra (Fig. S2 [31]) taken for a Li deintercalated film evidenced the end composition of Li$_{0.45}$NbO$_2$ [23]. Note that $1/eR_H$ were always smaller than nominal carrier concentrations estimated from $x$, as often observed in strongly correlated systems [1–4,32,33]. This discrepancy can be attributed to anomalous electronic states of this system as we will discuss later.

### B. Systematic transition from band insulator to superconductor associated with non-Fermi liquid behavior

Figure 2(b) shows the temperature dependence of resistivity ($\rho$) for Li$_{1-x}$NbO$_2$ films. At the lowest $1/eR_H$

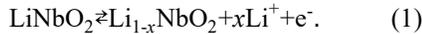

*Contact author: soma@mct.isct.ac.jp

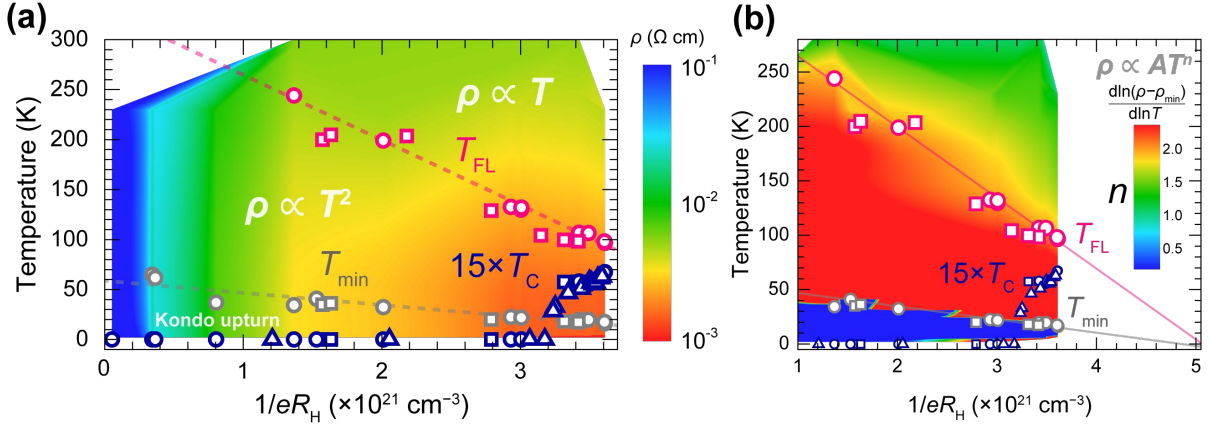

FIG. 3. Electronic phase diagram and quantum critical behaviors in $Li_{1-x}NbO_2$. (a) Temperatures that indicate transitions from NFL to FL ($T_{FL}$) states, Kondo upturn ($T_{min}$), and superconductivity ($T_c$) as a function of $1/eR_H$ determined from Hall-effect measurements. Circular, square, and triangular symbols correspond to the data obtained from three different devices (the correspondence is listed in Table S1). The contour plot represents the $1/eR_H$ dependent $\rho$–$T$ curves. Broken lines indicate linear fits to $T_{FL}$ and $T_{min}$. (b) The contour plot represents $d\ln(\rho-\rho_{min})/d(\ln T)$ being equal to temperature exponents $n$ ($\rho \propto T^n$).

$= 5.3 \times 10^{19}$ cm$^{-3}$, the film showed an insulating behavior ($d\rho/dT < 0$) at low temperatures, which was consistent with previous studies [18, 20–22]. With increasing $1/eR_H$, $\rho$ monotonically decreased and exhibited metallic behaviors ($d\rho/dT > 0$). In addition, SC appeared at low temperatures when $1/eR_H$ exceeded $3.0 \times 10^{21}$ cm$^{-3}$, and superconducting transition temperatures ($T_c$) increased with $1/eR_H$. Systematic increase in $T_c$ up to 4.5 K was clearly seen in a range of $3.1$–$3.6 \times 10^{21}$ cm$^{-3}$ [Fig. 2(c)]. Note that $T_c$ are defined by temperatures where $\rho$ reach 95% of the normal state $\rho$ at 5 K. The gradual drops in $\rho$ obeyed the Halperin-Nelson equation (Fig. S3 [31]) and thus evidenced the 2D SC [33]. The ideal and reversible Li-ion electrochemical reactions were clearly demonstrated by repeatable switching of SC (Fig. S4 [31]).

The normal state $\rho$–$T$ curves were further analyzed. In lightly hole-doped metallic states ($1/eR_H \leq 2.0 \times 10^{21}$ cm$^{-3}$), $\rho$ obeyed the $T^2$ dependence [Fig. 2(d)], reflecting Fermi-liquid (FL) behaviors [35]. In heavily hole-doped ones (i.e., $1/eR_H = 3.6 \times 10^{21}$ cm$^{-3}$), the $T$-linear dependence appeared instead of the FL behavior above ~100 K [Fig. 2(e)], suggesting a non-Fermi liquid (NFL) behavior associated with the 2D antiferromagnetic fluctuation [1,36]. The fact that the NFL behavior coincided the emergence of SC is reminiscent of strange metals associating with unconventional SC [1,2]. We define FL-NFL transition temperatures ($T_{FL}$) as temperatures at which $d(\rho-\rho_{min})/d(\ln T) = 2$ ($\rho-\rho_{min} \propto T^2$, where $\rho_{min}$ are minimal $\rho$) [8,12,14]. Indeed, $T_c$ were found to vary with $T_{FL}$ (Fig. S5 [31]), suggesting that the FL instability governed SC as often discussed in unconventional superconductors [1,37]. Moreover, upturns in $\rho$ seen at low temperatures were commonly characterized by the logarithmic dependence and tendency approaching to constants at the lowest temperature [Fig. 2(f)]. It is noteworthy that temperatures at $\rho_{min}$ ($T_{min}$) monotonically decreased with increasing $1/eR_H$. These behaviors were different from usual localization processes, and can be attributed to Kondo scattering (Fig. S6 [31]) [3,4,38–40].

The consideration of Kondo scattering needs some cautions since it usually arises from interactions between itinerant electrons and localized spins, which are unprecedented in $Li_{1-x}NbO_2$ [1,6]. Recent studies suggest that some flat-band systems are well described with Kondo lattices where some parts of bands are analogous to those of localized $f$-electrons in heavy fermion systems [4,13,14,28]. The observation of such a Kondo effect may also be the consequence of interplaying Nb $4d$ electrons if they play two roles of localized spins and itinerant electrons within a single $NbO_2$ layer to form Kondo singlets. This situation does not contradict the observed discrepancy between nominal $x$ and $1/eR_H$ and NFL behaviors of itinerant electrons [14,41]. The Kondo effects are associated with the enhancement of negative magnetoresistance (MR) below $T_{min}$, as we indeed observed in lightly hole-doped $Li_{1-x}NbO_2$ (Figs. S7A–C [31]). Therefore, the emergence of SC is closely related to magnetic fluctuations in 2D $NbO_2$ triangular lattice.

*Contact author: soma@mct.isct.ac.jp

## C. Electronic phase diagrams and quantum criticality

Having established the magnetotransport properties, we constructed electronic phase diagrams consisting of a bunch of $\rho$–$T$ curves and three characteristic temperatures ($T_{FL}$, $T_{min}$, and $T_c$) as a function of $1/eR_H$ [Fig. 3(a)]. The data taken from three devices with different film thickness are plotted together with different symbols (see Table S1 [31]). Similar dependencies of characteristic temperatures in all the devices guaranteed high reproducibility and homogeneity of Li-ion electrochemical reactions.

The SC appeared at $1/eR_H > 3.2 \times 10^{21}$ cm$^{-3}$ and $T_c$ seemed to trace a part of the dome-shaped dependence [Fig. 3(a)]. In addition, both $T_{FL}$ and $T_{min}$ showed the monotonic decrease toward the superconducting region, indicating similarities to characteristics of unconventional SC as quantum critical phenomena [1–4]. The FL instability is clearly captured in contour plots of temperature exponent $n = \mathrm{d}(\rho-\rho_{min})/\mathrm{d}(\ln T)$ [Fig. 3(b)] and coefficient $A = (1/2) \times \mathrm{d}^2\rho/\mathrm{d}T^2$ (Fig. S8 [31]). The decrease of $T_{FL}$ in a linear manner with $1/eR_H$ means the suppression of the FL state ($n = 2$) and evolution of the NFL state ($n < 2$). Note that $n$ became negative below $T_{min}$, reflecting the Kondo upturn. The characteristics of FL states can be further verified in the contour plot of $A$ as their magnitudes represent effective mass $m^*$ ($A \propto m^{*2}$) [1,3]. In the vicinity of the superconducting state, for example, $A$ coefficient was as large as $1 \times 10^{-7}$ Ω cm K$^{-2}$ and became comparable to those of heavy fermion systems [3,42,43]. We believe that strong magnetic fluctuation in NbO$_2$ triangular lattice leads to exceptionally strong correlation. Furthermore, extrapolating to $T = 0$, $T_{FL}$ and $T_{min}$ merged at $1/eR_H$ of $\sim 5 \times 10^{21}$ cm$^{-3}$, suggesting QCP. Given these correspondences and negative magnetoresistance (Fig. S7(d) [31]), it is naturally concluded that the superconducting dome is formed near the QCP by the magnetic fluctuation in the triangular lattice. It is worth mentioning that all $\rho$–$T$ curves are scaled when $\rho$ and $T$ are normalized by $\rho_{min}$ and $T_{min}$, respectively (Fig. S9 [31]). This result suggests that the concentration of itinerant carriers is a crucial parameter for controlling competition between Kondo singlets and Cooper pairs, being supportive to understand our electronic phase diagram in a unified manner.

## IV. DISCUSSION

The obtained phase diagram of Li$_{1-x}$NbO$_2$ becomes more general if it is illustrated in connection with known quantum critical systems [1–7] (Fig. 4). LiNbO$_2$ has a completely filled Nb 4$d_{z^2}$ single band and thus the non-magnetic state. In contrast, cuprates as typical hosts of quantum critical systems have half-filled Cu 3$d_{x^2-y^2}$ single bands that maximize the electron correlation in magnetic ordered states. The heavy hole doping to LiNbO$_2$ opens new route to access QCP starting from a trivial band insulator and reaching correlated metals. Interestingly, this approach is opposite to those for known systems where melting the initial magnetic ordered phases can be realized by hole doping and applications of pressure and magnetic field. Therefore, the phase diagram of Li$_{1-x}$NbO$_2$ should be reversely connected to typical one (cuprates in Fig. 4). In other words, our experiments traced phase boundaries in the *opposite* direction to cuprates. In this vein, hypothetical Li$_0$NbO$_2$ with a half-filled Nb 4$d_{z^2}^1$ band can be considered as another parent of superconducting Li$_{1-x}$NbO$_2$. Although irreversible Li-ion electrochemical reactions prevent us from accessing Li$_0$NbO$_2$ at the moment, it is suggested to be a Mott insulator by taking moderate $U$ in theoretical calculation [21]. In this scenario, SC appears in electron-doped Mott insulating 2$H$-NbO$_2$ [44].

*Contact author: soma@mct.isct.ac.jp

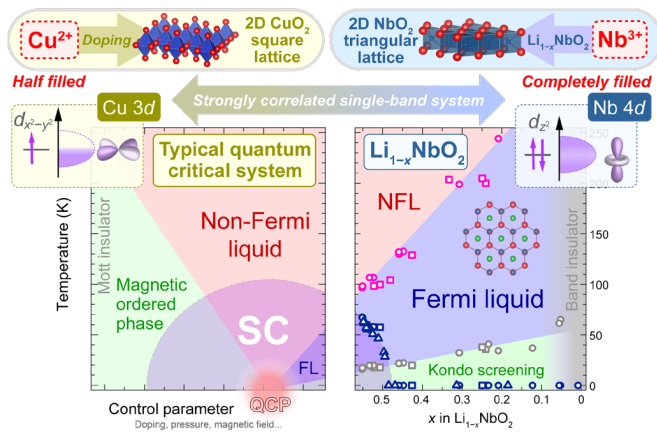

FIG. 4. Hypothesis on electronic phase diagram of Li$_{1-x}$NbO$_2$. The extended electronic phase diagram in a framework of quantum critical systems (1,3,5,6,7), represented by strongly correlated high-$T_c$ cuprates having a magnetic ordered phase with the half-filled Cu 3$d_{x^2-y^2}$ band. Upon hole doping (from right to left), the initial LiNbO$_2$ with the filled Nb 4$d_{z^2}$ band hypothetically undergoes to the *overdoped* region for cuprates to reach the superconducting dome under a NFL state. Then, Kondo screening in 2D NbO$_2$ triangular lattice may be present without wonder as often seen in the *overdoped* region for heavy-fermion systems. Note that $x$ in Li$_{1-x}$NbO$_2$ are not nominal values but those estimated from elemental and structural analyses, and a magnification factor of $T_c$ is 15.

Since SC in $Li_{1-x}NbO_2$ is evidently magnetically mediated, the magnitude of $T_c$ could reflect the strength of magnetic interactions [36]. In general, exchange coupling in early-transition-metal oxides with small hybridization is weaker than in late-transition-metal oxides [5,45]. Therefore, $T_c$ of $Li_{1-x}NbO_2$ is lower than cuprates as in the case of SC in heavy fermion compounds that are driven by magnetism of $f$ electrons [1,3,36]. It seems to obey the hypothesis that strong correlation is imperative for unconventional SC while $T_c$ are mainly governed by the strength of magnetic interactions. Nevertheless, unveiling phase diagram of an early-transition-metal oxide is significant because the emergence of superconducting dome near the magnetic QCP is again major finding for a strongly correlated superconductor.

Apart from general aspects, we would like to focus on unprecedented roles of spin-frustrated 2D triangular lattice of $NbO_2$ layers. The emergence of the Kondo effect in the FL state contradicts naive picture on overdoped FL states with no magnetic interaction [1,5]. However, recent studies suggested that it likely happens in cuprates and Moiré graphene [46,47]. As for $Li_{1-x}NbO_2$, the Doniach picture of heavy-fermion systems is rather appropriate as discussed already [1,6,48]. In this picture, a spin-liquid state can be formed if localized spins are highly frustrated [49]. Since the hypothetical host $Li_0NbO_2$ has a 2D triangular lattice with $S = 1/2$, the suppression of Kondo screening observed in $Li_{1-x}NbO_2$ manifests itself in the vicinity of quantum spin-liquid state [3,49], which is sometimes suggested to trigger exotic SC [50]. In fact, quantum spin liquid behaviors and the emergence of the superconducting dome with NFL behaviors has been discovered in 2D organic compounds with triangular lattices [51,52]. Modulating Kondo effect that relieves frustration within $NbO_2$ layers would have led to the formation of Cooper pairs. Thus, $Li_{1-x}NbO_2$ should not be classified simply as a layered oxide, but also as a class of frustrated flat-band systems and/or heavy-fermion compounds. This perspective provides new insights that pave the way for a general understanding of various strongly correlated superconductors.

## V. CONCLUSIONS

In summary, we have revealed the electronic phase diagram of the layered superconductor $Li_{1-x}NbO_2$ using the Li-ion electrochemical cells combined with epitaxial films. As a result of fine hole-doping steps, we demonstrated the superconducting dome associated with the quantum criticality in 2D and spin-frustrated $NbO_2$ lattices. This discovery contributes to a deeper understanding of strongly correlated superconductors.

## ACKNOWLEDGMENTS

This work was supported by JST PRESTO (JPMJPR22Q3), JSPS KAKENHI (JP20K15169, JP21H02026, JP22H04505, JP24H01179, JP24H00480), Yoshinori Ohsumi Fund for Fundamental Research, and The Foundation for The Promotion of Ion Engineering.

*Contact author: soma@mct.isct.ac.jp

*Contact author: soma@mct.isct.ac.jp